\RequirePackage[l2tabu,orthodox]{nag}
\documentclass[11pt,letterpaper]{article}

\usepackage[T1]{fontenc}
\usepackage[utf8]{inputenc}
\usepackage{ebgaramond}
\usepackage{tgheros}
\usepackage[strict,autostyle]{csquotes}
\usepackage[USenglish]{babel}
\usepackage{microtype}
\usepackage{authblk}
\usepackage{datetime}
\usepackage{geometry}
\usepackage{hyperref}
\usepackage{setspace}
\usepackage{titlesec}
\usepackage{url}

\AtBeginEnvironment{quote}{\singlespacing\small}

\titleformat{\section}{\normalfont\sffamily\large\bfseries}{\thesection.}{0.3em}{}

\newdateformat{monthyeardate}{\monthname[\THEMONTH] \THEYEAR}

\newcommand{\myname}{Geoff Boeing}

\newcommand{\myaffiliation}{Department of Urban Planning and Spatial Analysis\\University of Southern California}
\newcommand{\papertitle}{We Live in a Motorized Civilization: Robert Moses Replies to Robert Caro}
\newcommand{\paperkeywords}{History, Robert Moses, Transportation, Urban Planning}

\hypersetup{
	pdfauthor={\myname},
	pdftitle={\papertitle},
	pdfsubject={\papertitle},
	pdfkeywords={\paperkeywords},
	pdffitwindow=true,
	breaklinks=true,
	colorlinks=false,
	pdfborder={0 0 0}
}

\begin{document}

\title{\papertitle}
\author[]{\myname}
\affil[]{\myaffiliation}
\date{\monthyeardate\today}

\maketitle

\bigskip

\section*{Introduction}

In 1974, Alfred A.\ Knopf, Inc.\ published Robert Caro's \textit{The Power Broker}, a critical biography of Robert Moses's dictatorial tenure as the \enquote{master builder} of mid-century New York. Moses profoundly transformed New York's urban fabric and transportation system, producing the Brooklyn Battery Tunnel, the Verrazano Narrows Bridge, the Westside Highway, the Cross-Bronx Expressway, the Lincoln Center, the UN headquarters, Shea Stadium, Jones Beach State Park, and many other projects. However, \textit{The Power Broker} did lasting damage to his public image and today he remains one of the most controversial figures in city planning history. Caro's book won the 1975 Pulitzer Prize and was eventually named one of the Modern Library's 100 greatest nonfiction books of the twentieth century. Prior to this critical acclaim, however, came a pointed response from the biography's subject.

On August 26, 1974, Moses issued a turgid 23-page statement denouncing Caro's work as \enquote{full of mistakes, unsupported charges, nasty baseless personalities, and random haymakers.} In his statement, Moses denied having had any responsibility for mass transit, dismissed his \enquote{lady critics,} and defended the forced displacement of the poor as a necessary component of urban revitalization: \enquote{I raise my stein to the builder who can remove ghettos without moving people as I hail the chef who can make omelets without breaking eggs.} The statement also included a striking passage on what Moses perceived to be his critics' hypocrisies in the modern age of automobility: \enquote{One cannot help being amused by my friends among the media who shout for rails and inveigh against rubber but admit that they live in the suburbs and that their wives are absolutely dependent on motor cars. We live in a motorized civilization.}

In the wake of Moses's statement, Caro issued a (much shorter) same-day reply. The following morning, \textit{The New York Times} published an article reporting on their testy exchange.\footnote{Kaufman, Michael T. 1974. \enquote{Moses Rips Into \enquote{Venomous} Biography.} \textit{The New York Times}, p. 25, August 27, 1974.} Moses's original typewritten statement survives today as a grainy photocopy in the New York City Parks Department archive. To better preserve and disseminate it, I have extracted and transcribed its text using optical character recognition and edited the result to correct errors. In the pages that follow, I have compiled my transcription\footnote{Available at \href{https://github.com/gboeing/moses-caro}{https://github.com/gboeing/moses-caro}} of Moses's statement, followed by Caro's reply to it.

\newpage

\section*{Comment on a New Yorker Profile and Biography}

\noindent Robert Moses\\
\noindent August 26, 1974\\
\vspace{1cm}

\noindent Robert A. Caro has written a biography about me, excerpts from which have appeared in a four-part Profile in \textit{The New Yorker.} The Profile seems to be a considerably digested, amended, edited and expurgated version of the twelve hundred page biography.

I have sought the advice of close friends, most of whom told me to say nothing. On the other hand, judging by inquiries, some comment seems to be called for. I have limited it to less than 3500 words as against 120,000 in the Profile and 600,000 in the book. I shall have nothing further to add, except to suggest that the critics remember Governor Smith's favorite remark in such contexts, \enquote{Let's look at the record.}

It is difficult to address oneself at once to a whitewashed, Bowdlerized Profile with a sensational, eye-catching title calculated to attract curious readers, and a biography. The Profile is within bounds of current journalistic practice. The biography, on the other hand, is full of mistakes, unsupported charges, nasty, baseless personalities and random haymakers thrown at just about everybody in public life.

The biography tries to prove that I was a good boy who fell from grace, became a politician and mistreated the poor. The author can't even get the names and places straight. There are hundreds of careless errors.

Many charges are downright lies, manufactured by early opponents who have waited for an opportunity to satisfy ancient grudges and found this author a ready instrument. Among personal, nasty, false, venomous and vindictive canards is one that I was romantically linked with Mrs.\ Ruth Pratt, our first Congresswoman, and that as a result my wife took sick, became an alcoholic and a recluse, that I virtually abandoned her, joined some kind of foursome and took a lady who subsequently became my wife to Florida. My companion on the short Florida vacations was my daughter Jane. With respect to my family life, the author's innuendos are wholly untrue and scurrilous. Under the fair libel laws they would be actionable.

For dirt and misinformation the author says he latched onto a brother of mine, now dead. Similarly, he attacks by name, by sly hints of wrongdoing and conflicts of interest scores of prominent and respected officials and citizens, many of them dead.

The author praises two reporters, Gleason and Cook of the Scripps-Howard press, as sterling representatives of the journalistic profession, but fails to inform his readers that they were called in by the district attorney and fired in disgrace after being forced to admit that they had fabricated a particularly vicious housing story.

I invite no prolonged controversy with the likes of Caro and his publishers. This comment is not meant to spark controversy. It aims to make one statement which will answer legitimate inquiries. If there were the slightest vestige of truth in the random charge that poor, helpless, displaced persons met ruthless public works dictators who sadistically scattered them to the worst rookeries, why do not Caro and his publisher offer some plausible evidence? Ninety-eight percent of the ghetto folks we moved were given immeasurably better living places at unprecedented cost. Usually a month after the last relocation not a letter of complaint was received.

And what in ordinary English is the meaning of the words \enquote{Power Broker?} What is a power broker? Does Fleischman of \textit{The New Yorker} or Knopf, the publisher, or Caro, the snooper, mean that I profited financially? If that's the implication why don't they say so? Do they mean I have been what in common political parlance is called an \enquote{operator?} And what is an operator? Someone who is devious, serpentine, anfractuous? Do these publishers use such words simply to attract gullible purchasers regardless of what there is between the covers?

As to whether honest public service is a profitable profession, let me say that I spent most of my principal in order to remain in public employment. The story of my wealth is fiction. I had to borrow money from my mother to persuade busted contractors to bring back material in order to open Jones Beach on time. I had to dun friends to keep survey parties from starving when enemies cut off our funds in the State Legislature. These tidbits escaped the author, Knopf and Fleischman.

There is no reliable evidence to be obtained from a few landowner malcontents who profited less than they expected from our improvements. To find out whether there were any sizable numbers of families displaced to accommodate parks, parkways, bridges, tunnels, power developments and suchlike, it would be necessary to make a thorough impartial canvass, not to interview a few bellyachers at street corners or disgruntled truck farmers on the edge of the City about to make hundreds of thousands of dollars from proximity to new roads.

It would also be essential to determine whether any growth or prosperity on Long Island east of the City line would have been possible without the facilities for travel we built and what could have been substituted for cars, trucks and buses running on rubber and paved roads. One can not help being amused by my friends among the media who shout for rails and inveigh against rubber but admit that they live in the suburbs and that their wives are absolutely dependent on motor cars. We live in a motorized civilization.

The current fiction is that any overnight ersatz bagel and lox boardwalk merchant, any down to earth commentator or barfly, any busy housewife who gets her expertise from newspapers, television, radio and telephone, is ipso facto endowed to plan in detail a huge metropolitan arterial complex good for a century. In the absence of prompt decisions by experts, no work, no payrolls, no arts, parks, no nothing will move. Honest public officials will be denounced as wheelers and dealers, oppressors of the poor, dictators, fixers and power brokers intolerable in a true democracy. Officials who are not thin skinned and have the courage of their convictions pay little mind to the gravamen or such charges and don't bother to answer allegations at length or defy all allegators.

What with poisonous critics, savage commentators, public relations advisers, speech and ghost writers and equal time rebuttal spouters over the air, we are rapidly succumbing to what Whitney Griswold of Yale used to aptly call \enquote{nothing but technological illiteracy.} On the other hand, you don't catch us minuscule, imitation Napoleons denying open forums for dissidents to find fault and enthusiasts to advocate causes. We simply ask that the forums be properly conducted. Other victims of biographies have survived harder impeachments. Critics claim to be anxious over the influence of petty dictators dres't in a little brief authority. On the other hand, builders worry about government paralyzed by lunatic fringes.

In this context I think of that kindly fuddy duddy, British Poet Laureate Sir William Watson, who was much too decent for the critics. He was roused to fury only once. That was when he could no longer stand the gibes of the wife of the Prime Minister and burst out with the best verse he ever wrote, beginning:

\begin{quote}
\enquote{She is not old, she is not young,\\
The Woman with the serpent's tongue.}
\end{quote}

Here and there in the Profiles there are broad hints that my associates and I were not always ultra refined in our actions. They complain that we have not followed the Marquis of Queensberry rules. They say that on occasion we quietly after hours smoothed the paths for our parkways. They insinuate that old trees were whisked away by ingenious stump pullers to allay the apprehensions of nervous environmentalists. If this be true, tell it not in Gath. Publish it not in the streets of Askelon. As the city folk ride into the open country we shall, I trust, be forgiven. The original railroad builders too were in a sense fuel merchants and chopped down some spindly woods to stoke their engines.

By way of contrast I like to reflect on the approach of one of our own distinguished American philosophers and all-around tongue and bat athletes, Leo Durocher. Perhaps in mild extenuation of his own occasional lapses from Amy Vanderbilt's rules of etiquette, Leo wisecracked, \enquote{Nice guys finish last.} It is well in this context to remember the nameless, forgotten pitcher in Casey at the Bat, the anonymous hero who struck Casey out.

The parks and rights-of-way we snatched just ahead of the sales of big estates to farsighted realtors bent on subdivision on the most profitable terms permitted by pliable zoning commissions would cost from ten to twenty times as much today as vacant land, and where it was necessary to take houses, probably a hundred times. The critics and second guessers say we were sometimes rude, arbitrary and highhanded. Maybe so, but suppose we had waited. Critics are ex post facto prophets who can tell how everything should have been done at a time when they were in diapers, in rompers or invisible. We ditchdiggers do our best to live up to our oaths of office with the slender talents vouchsafed us. We enjoy our work. We accept its drawbacks without whining. We expect neither full understanding by the Caros and Knopfs, nor unqualified popular acclaim by \textit{The New Yorker.}

We don't ask anyone to be sorry for us. There is considerable confetti among the brickbats. A park commissioner must have a cold heart on a balmy day at Jones Beach not too pleased to be recognized by a few Beach aficionados, or perhaps in the Central Mall restaurant to get a friendly glance of the eye and even a wet smack from one of those nice middle-aged ladies who fancy they owe him something. Perhaps these occasional harmless little compliments represent only mistaken identity, but do the beneficiaries of our parks ever stop to ask the publishers for their autographs?

Charges of arrogance, contempt for the so-called democratic process, lack of faith in the plain people, brutal uprooting and scattering of those in the way are as old as recorded history. In such periods the left wingers, fanatical environmentalists and seasonal Walden Ponders have a field day. They believe that Steve Ben\'et's termites, who eat steel columns and beams, will soon level the tall buildings and bridges of every metropolis and enable us to retire to unspoiled and untrammeled nature.

Anyone in public works is bound to be a target for charges of arbitrary administration and power broking leveled by critics who never had responsibility for building anything. I raise my stein to the builder who can remove ghettos without moving people as I hail the chef who can make omelets without breaking eggs. Those of us who engage in the dangerous trade of public works expect such pot shots and, short of libel, take them good-naturedly. This reminds me of the remark attributed to a departed statesman: \enquote{Enemies---I have no enemies. I buried all those bastards long ago.} Any administrator the critics don't like is a small scale Hitler, Mussolini or Stalin.

The author's thesis is that I was once a pilgrim who made progress, fell among charlatans, lost his inspiration and never reached the Celestial City. This is a sad summary to which most observers on reflection are not likely to subscribe. I prefer to believe that some of us may still be in demand when politicos of greater refinement and more sweetness and light have been found wanting.

The Caros think that those who have found it necessary to expand their activities to meet the responsibilities imposed on them do so out of vanity and the yen to collect titles, hats, badges, medals and diplomas as though they were Phi Beta Kappa keys of departed scholars for sale in Bowery pawnshops. To be sure, there are sensitive folk who have suffered from official arrogance and dislike what the military call the habit of command. The medieval ideal was of course Chaucer's improbable verray, parfit gentil knight sans peur et sans reproche, a breed which survives only in museum tapestries.

The author of much of this diatribe has no remote notion of how commissioners are made. I like to recall riding in an elevator with several other officials to a meeting at the City Comptroller's office. As we left the car a chum of the elevator man asked him, \enquote{Who are they?} The operator replied, \enquote{Commissioners. A dime a dozen.} The author knows very well that most commissioners usually rise to the top by doing their menial chores modestly, have the right political sponsors, are loyal, clean the windows, sweep the floor and polish up the handle of the big front door. They polish it up so carefully that they become the rulers of the Queen's Navee.

To be sure, the builder is primarily concerned with bringing home the bacon. Ultra-refined, finicky folk won't eat oxtail ragout because of its lowly associations. Too much refinement can paralyze engineering. Sometimes it seems sheer madness to enter such a thankless profession, but it has its compensations. In spite of these drawbacks it is astonishing how many recruits leap to their feet as the bugler shatters the dawn with the first notes of reveille and they hear the irresistible clarion call to public service and sacrifice.

The author and publisher do not comprehend the obligations of leadership. It is true that altogether too much limelight falls on the so-called stars as compared to the rest of the cast and the workers behind the scenes. Esprit de corps, more commonly called teamwork, is the most important factor in leadership and this can not be asserted. It must be earned. It is not high aim, or courage, knowledge, wisdom, nor even the fighting spirit that keeps enterprises of great pith from going astray in spite of treachery, treasons, stratagems and spoils. It is personal, never-failing loyalty, not loyalty to abstract principles, but unshakeable personal, never-failing loyalty which gives support in the clinches. The greatest satisfaction is to find you have loyal friends. The others are not worth a tinker's damn.

The author minimizes our recapture and reclamation of the neglected New York waterfront, an accomplishment without parallel in any other great port city in the world. He accepted as gospel yarns spun by Elwood M Rabenold, once a West Side legislator. Rabenold boasted a Harvard accent superimposed on Pennsylvania Dutch. He high-hatted and thereby won the undying enmity of Jimmy Walker, the Senate leader who represented the next district. Rabenold seems to have disliked me because as head of the State park system I charged him with being mixed up in Palisades real estate wheeling and dealing. Caro unearthed Senator Rabenold who by that time had retired to a huge, biblical Pennsylvania farm with herds of cattle on a hundred hillsides and was leader of the Lutheran Church. I told Caro this man had been sent to Sing Sing for three years for stealing as counsel from an incompetent old lady and had been disbarred from the practice of law in New York. This author says Walker railroaded him but does not say where or how. Walker had nothing whatever to do with it, but Caro seems still to regard Rabenold as a reliable source of information.

A magnificent car the author refers to was my mother's huge old Marmon, called by her driver a Mormon. After her death I used it in the Long Island State Park system, when I gave it to the Commission. It was worth under \$1,000. Similarly, one of the yachts was a ram runner equipped with airplane engines bought at a Sherriff's sale for \$300. It caught fire on a trial trip, sank and was then equipped with heavy duty engines and served the Commission some twenty years. Another magnificent luxury boat was bought by me for less than a thousand dollars. I ran it. The magnificent meals dreamed up by the author were cooked on sternos. My entire City Trust Company banking report as Moreland Commissioner under Roosevelt and Lehman was written longhand in a Western Union tower on Fire Island given me by the head of Western Union, furnished by me, given the State and lost in the 1938 hurricane. The best boat the park authorities had was bought from Charles P. Noyes, the realtor, for one-third its value when he was selling the effects of one of his boys killed in a car accident. There are countless such evidences in Caro's book of his exuberant Oriental fantasies.

The stink bombs of some lady critics don't suffocate us. Several of these characters said there was not one note of beauty and no vestige of good taste or culture at both World's Fairs at Flushing Meadow and that both were offensive and a total loss. The huge task of reclaiming this fetid meadow blocked by the biggest ash dump in municipal history, so well described in \textit{The Great Gatsby}, and widening a foul, tortuous, muddy brook into huge, beautiful lakes, all this was ignored by these ladies. I allude to those to whom I refer. Wild horses would not drag from me the names of these representatives of the not always fair sex.

I find most critics, male and female, enormously diverting as they decide at lunch or cocktail: which curtains to lower and shows to close, which actors to get the hook and which public officials to send to the hoosegow for getting their fishhooks caught in the cracker jar. These critics distribute with equal justice and impartiality the crowns of wild olive and the kisses of death.

We badly need another William Cowper and another George Crabbe in our approach to the press and television. They strove more than two centuries ago and couched their indictments in forthright Anglo Saxon. Cowper was the tougher of the two. Crabbe was aptly named. He learned about human nature by starting life as an ordinary midwife and then got into Socratic midwifery and became a great writer, cleric and chronicler of the age. Let me give you a claw of Crabbe:

\begin{quote}
\enquote{I sing of News, and all those vapid sheets\\
The rattling hawker vends through gaping streets;\\
Whate'er their name, whete'er the time they fly,\\
Damp from the press, to charm the reader's eye:\\
For, soon as morning dawns with roseate hue,\\
The Herald of the morn arises too;\\
Post after Post succeeds, and, all day long,\\
Gazettes and Ledgers swarm, a noisy throng.}
\end{quote}

The author does not have the remotest notion of the independent corporate character of a World's Fair and confuses it with public business. He is even more ignorant of the laws and practices of independent public authorities. In public authority financing the entrepreneur must sell its bonds on character, not on government guarantees against loss. Any biography which rests on any other assumption is irresponsible and those who aid and abet in publishing it are unreliable. The nasty Caro-Knopf charges about patronage and favored banks and other institutions are made out of whole cloth, were never sustained and even repudiated by the courts.

Caro asserts boldly, and of course without any credible proof, that I took funds which could have been used to build mass transit facilities, and used them to build parks, parkways and roads. The absurdity of such a statement is that there never were any such funds available or, in any event, under my control. I never had anything to do with the building of such facilities in any official or other capacity. In short, I never was in charge of mass transit.

Various people, politicians and amateur strategists are repeatedly saying more should be done for mass transit. I agree. But why don't these people do something about it instead of talking? Why don't they secure the funds and build them as I did in the areas I was charged officially? As to selection of sites for housing, each one we built was approved by the Board of Estimate and other bodies after long processing and, in some cases, strenuous local opposition.

Similarly as to parks, parkways, bridges and other facilities, most of them encountered serious opposition, and all were built, after official government approval, and in New York City with the consent of the Board of Estimate.

On the personal side, Caro says, proving I don't know what, that I am a bum speaker, rushing through my text without stopping for applause and contemptuous of the audience. The little weasels who charge conflict of interest think a Christmas present of a bottle of old brandy from a contractor calls for returning it publicly with a big flourish as an attempted bribe. The Caros always look for what the bootlegger in Scott Fitzgerald's Gatsby called \enquote{gonnections.}

The author's thesis, as I have said, is that beginning as an idealistic reformer I became a power broker and patronage dispenser and ended up a bitter, irascible, disillusioned, arthritic old curmudgeon dependent on thick eye glasses and a big hearing aid and sporting huge liver spots and a butler's pantry or bulge in the midriff, a resemblance which will be greeted with derisive whoops and Bronx cheers from my buddies at the beaches, bars and bistros.

Caro denounces just about everybody. Like Jeremiah, he finds no balm in Gilead and, like Nathanael,\footnote{The name \enquote{John the Baptist} appeared here in the typewritten original, but was crossed out and \enquote{Nathanael} was handwritten above it. Nathanael is the correct reference to John 1:46.} questions whether any good thing ever came out of Nazareth. It may well be that \textit{The New Yorker} can sell its Profiles for two bucks, but I doubt that many well-heeled readers will fork out \$17.95, plus sachet, to read the unexpurgated Caro.

In appraising the qualifications of a writer to become an authority on public works there is really no substitute for successful experience and results visible to the naked eye. A foundation fellowship is no credential. It may only represent a poor investment and the familiar triumph of hope over experience.

Caro's engineering and transportation outgivings are ridiculously amateurish, naive and infantile. He picked them up from a disgruntled young engineer with the City Planning Commission who indulged in nasty recriminations after he left the City Planning Commission. Caro's idea was that if rapid transit rails were put in the middle of the Long Island Expressway which he didn't like, the entire problem would be solved.

I never called Mayor LaGuardia any of the so-called names Caro maliciously mentions and nobody will substantiate these miserable yarns. I did once say facetiously when the Sainted Fiorello put on a highly dramatic performance that he reminded me of Rigoletto. I am sorry I said it, although it was entirely innocent. I thought then and still do that he was the best mayor in my time.

Better judges than the Caros and Knopfs have ridiculed the practitioners of public works. In this context I always tip my hat to my cousin Frank Lloyd Wright. In his wilder projects of Welsh fantasy he really believed his own architectural interruptions at nature lifted us to the hills whence cometh our light, enhanced the plains and swept us out to the limitless sees. Frank's comparison of himself as a skylark and to me as a blind night crawler were of course just a quaint bit of Celtic humor. We take these things from the Frank Lloyd Wrights because we admire them in spite of their idiosyncrasies.

Those of us who have comparatively little time left for constructive work and have to husband their resources can not afford to waste muscle on keyhole snoopers, dirt dishers, gossips, and embittered bums trying to get hunk on someone they dislike.

In all patriotic yarns there is supposed to be a hero like Arnold von Winkelried who gathers the spears of the enemy to his bosom and saves the Swiss Navy. The role is not for me. The spears in this instance are made of the paper unelegently known as bumwad. They crumble on compact.

\textit{The New Yorker} Profile series ends with these remarkable sentences: \enquote{It is impossible to say that New York would be a better city if Robert Moses had not shaped it. It is possible to say only that it would be a different city.} Assuming that the City changes, however brought about, were as extensive as the author says, what would New York look like without them? Surely it could not have been left in a powerless and brokerless state of chassis and suspended animation.

The author says he is about to do a biography of Fiorello LaGuardia. I suggest to Marie LaGuardia that she be very careful.

\newpage

\section*{Robert Caro's Reply to Robert Moses}

\noindent Robert A. Caro\\
\noindent Courtesy of Random House\\
\noindent August 26, 1974
\vspace{1cm}

\noindent One aspect of Robert Moses that my book attempts to portray is that of the smearer of reputations, the purveyor of baseless innuendo and outright falsehood, the wholesaler of defamation---in a public official who, through a ruthless and awesome public relations machinery, practiced McCarthyism long before there was a McCarthy. I am not displeased that the Commissioner has furnished this additional up-to-date documentation of all this and would be pleased to let my book speak for itself were it not for his statements and attempts to drag others into the picture.

I cannot, for example, let pass his attempt to destroy his brother's reputation in death as he did in life. As my book documents, no fewer than six separate officials of the La Guardia administration are aware of how Robert Moses hounded Paul Moses, a brilliant, competent, respected engineer, out of public service and forced him down the road that lead him to live out the last thirty years of his life in the most abject and humiliating poverty.

As for the story of Robert Moses and his brother's inheritance, anyone can find the truth of what I wrote simply by examining the records and affidavits of the surrogate's court and New York county supreme court, the liber numbers of which can be found in the source notes of my book, page 1210. And I wish the Commissioner would stop quoting Al Smith, a unique, great and wonderful man, in an attempt to bolster his own broad scale billingsgate.

Mr.\ Moses asks, \enquote{What is a power broker?} If the Commissioner tells me which of those two words he doesn't understand, I will be glad to point it out to him in a dictionary. A \enquote{power broker} deals in power as a real estate broker deals in real estate---and Robert Moses has been the supreme dealer in power in New York City and New York State for almost half a century.

It is slightly absurd (but typical of Robert Moses) to label as without documentation a book that has 83 solid pages of single-spaced, small-type notes and that is based on seven years of research, including 522 separate interviews.

\end{document}